\documentstyle[11pt]{article}
\begin{document}
{\setlength{\oddsidemargin}{1.6in}
\setlength{\evensidemargin}{1.6in}
}
\baselineskip 0.55cm
\title{On the variable-charged black holes in General
Relativity: Hawking's radiation and naked singularities}

\author
{Ng. Ibohal,\\
Dept of Mathematics, Manipur University,\\
Imphal 795003, Manipur, India\\
E-mail: ngibohal@rediffmail.com}
\date{}
\maketitle

\begin{abstract}
        In this paper the {\sl variable}-charged non-rotating Reissner-Nordstrom
as well as rotating Kerr-Newman black holes are discussed. ``Such
a variable charge $e$ with respect to the polar coordinate $r$ in
the field equations is referred as an electrical radiation of the
black hole''. It is shown that every electrical radiation $e(r)$
of the non-rotating black hole leads to a reduction of its mass
$M$ by some quantity. If one considers such electrical radiation
to take place continuously one after another for a long time, then
a continuous reduction of the mass may take place in the black
hole body and the original mass of the black hole may be
evaporated completely. At that stage, the gravity of the object
may depend only on the electromagnetic field, not on the mass.
Just after the complete evaporation of the mass, if the next
radiation continues, there may be a creation of a new mass leading
to the formation of {\sl negative mass naked singularity}. It
appears that this new mass of the naked singularity would never
decrease, but might increase gradually as the radiation continues
forever. A similar investigation is also discussed in the case of
{\sl variable}-charged rotating Kerr-Newman black hole. It has
been shown by incorporating of {\sl Hawking's evaporation} of
radiating black holes in the form of space-time metrics, every
electrical radiation of {\sl variable}-charged rotating as well as
non-rotating black holes may produce a change in the mass of
the body without affecting the Maxwell scalar.        \\\\\
\end{abstract}

\begin{center}
{\bf 1. Introduction}
\end{center}
\setcounter{equation}{0}
\renewcommand{\theequation}{1.\arabic{equation}}

        In General Relativity, charged black holes are defined as solutions of
Einstein-Maxwell field equations. So, it is well-known that the
Kerr-Newman solution is a rotating charged black-hole with three
constant parameters $M$, $e$, $a$ of the family; $M$ representing
the gravitational mass, $aM$ the angular momentum and $e$ the
charge of the body. When $a = 0$, the metric reduces to the
non-rotating Reissner-Nordstrom solution with the charge $e$. When
$e = 0$, the space-time metric reduces to the vacuum Kerr family
and when $a = e = 0$, the solution becomes the Schwarzschild
solution. The charged Kerr-Newman black hole has an {\sl external
event horizon} at $r_{+}=M+\surd {(M^2-a^2-e^2)}$\, and an {\sl
internal Cauchy horizon} at $r_{-}=M-\surd {(M^2-a^2-e^2)}$. The
{\sl stationary limit} surface $g_{tt}>0$ of the rotating black
hole i.\,e. $r=r_e(\theta)=M+\surd {(M^2-a^2{\rm
cos}^2\theta-e^2)}$ does not coincide with the event horizon at
$r_{+}$ thereby producing the {\sl ergosphere}. This stationary
limit coincides with the event horizon at the poles $\theta=0$ and
$\theta=\pi$ [1]. However, in the case of the non-rotating charged
Reissner-Nordstrom solution with $a=0$, the event horizon $r_{+}$
coincides with the stationary limit at $r_e$.

         The {\sl Hawking radiation} [2] suggests that black holes
which are formed by collapse, are not completely black, but emit
radiation with a thermal spectrum due to quantum effects [3]. That
objects radiate and thereby must decrease in size. Hawking [4]
stated --``Because this radiation carries away energy, the black
holes must presumably lose mass and eventually disappear. If one
tries to describe this process of black hole evaporation by a
classical space-time metric, there is inevitably a naked
singularity when the black hole disappears''. In an introductory
survey Hawking and Israel [5] have discussed the black hole
radiation in three possibilities with creative remarks --``So far
there is no good theoretical frame work with which to treat the
final stages of a black hole but there seem to be three
possibilities: (i) The black hole might disappear completely,
leaving just the thermal radiation that it emitted during its
evaporation. (ii) It might leave behind a non-radiating black hole
of about the Planck mass. (iii) The emission of energy might
continue indefinitely creating a negative mass naked
singularity''. Boulware [6] suggested that {\sl the radiation may
expressed in terms of the stress energy momentum tensor associated
with the field whose quanta are being radiated}. Thus, the change
in the metric due to the radiation may be calculated by using the
stress energy tensor in Einstein's field equations. If one were to
interpret $r$, $\theta$, $\phi$ as representing spherical polar
coordinates, there is a singularity at the origin $r=0$, whereas
when $r = \infty$, the metric approaches the Minkowski flat
metric. So the nature of the black holes depends on the polar
radius $r$. The question arises what happens when the charge $e$
is considered to be a function of $r$ in the stress energy
momentum tensor $T_{ab}$ of the Einstein-Maxwell field equations
of the rotating as well as non-rotating charged black holes. Here
it is to mention that Vaidya [7] could produce a metric describing
a radiative black hole by considering the mass of the
Schwarszchild solution, variable with respect to the retarded time
$u$.

        In this paper we try to incorporate the Hawking radiation
effects by considering Boulware's suggestion that the
energy-momentum tensors of electromagnetic field must have
different forms from those of Reissner-Nordstrom as well as
Kerr-Newman black holes as these two black holes seem not to have
any direct Hawking radiation effects. That is, the only idea left
is to consider the charge $e$ to be variable with respect to the
coordinate $r$ of these black holes. Here, we mean the {\sl
`electrical radiation'} of a charged black hole as {\sl the
variation of the charge $e$ of the body with respect to the
coordinate $r$} in the stress energy momentum tensor of the
Einstein-Maxwell field equations. So, for every electrical
radiation we consider the charge $e$ to be a function of $r$ in
solving the Einstein-Maxwell field equations and we have shown
{\sl mathematically} how the electrical radiation induces to
produce the changes of the mass of {\sl variable}-charged black
holes. One may incorporate the idea of losing (or changing) mass
at the rate as the electrical energy is radiated from the charged
black hole. It is noted that these results, presented in this
paper are mainly {\sl mathematical}. We do not, therefore intend
to determine the exact change of the mass numerically, but could
certainly observe mathematical result of losing (or changing) mass
in the space-time metrics cited below after every electrical
radiation.

        If the energy momentum tensor is of electromagnetic fields with the
charge $e(r)$, the Einstein-Maxwell field equations for both rotating and
non-rotating charged black holes yield the Ricci scalar $R\,(= R_{ab}\,g^{ab})$
not equal to zero. But for electromagnetic fields this Ricci scalar has to
vanish in general. As this Ricci scalar $R$=$R_{ab}\,g^{ab}$ for electromagnetic
fields vanishes, one would find that some quantity $m_1$ is being decreased from
the original masses of the black holes. However, it is found that the form of
the Maxwell scalar $\phi_1={1\over 2}
F_{ab}(\ell^a\,n^b+\overline{m}^a\,m^b)$
of the black holes remains unchanged.

        Thus, one might summarize the result of the paper in the form of
theorems:

{\bf Theorem 1.} {\sl Every electrical radiation of
`variable-charged' non-rotating Reissner-Nordstrom and rotating
Kerr-Newman black holes may produce
a change in the mass of the bodies without affecting the Maxwell
scalar}.

{\bf Theorem 2.} {\sl During the radiation process, after the
complete evaporation of masses of both `variable-charged'
non-rotating Reissner-Nordstrom and rotating Kerr-Newman
black holes, the electrical radiation may continue indefinitely
creating negative mass naked singularities}.

It appears that this theorem 2 may be in favor of
Hawking-Israel third possibility quoted above, but a violation
of Penrose's cosmic-censorship hypothesis that no naked
singularity can never be created [1]. The theorem 2 has been
presented in the form of classical space-time metrics below.

   Here, we use the word {\sl `change in the mass'} rather then
{\sl `loss of mass'} as there may be  possible of creation of mass
after the exhaustion of the original mass, if one repeats the same
process of electrical radiation. This may be seen latter in this
paper. Hawking radiation is being incorporated, in the classical
General Relativity describing the change in mass appearing in the
classical space-time metrics, without quantum mechanical aspect as
done by Hawking [2] or the path integral method used by Hurtle and
Hawking [8] or thermodynamic viewpoint [9]. We present classical
space-time metrics affected by the  change in the mass of the
variable charged black holes after electrical radiation in Section
2. In Section 3 the properties of the metrics formed after the
electrical radiation is discussed in regards with Kerr-Schild form
and Chandrasekhar's relation. The NP version of original
Reissner-Nordstrom and Kerr-Newman solutions are presented in an
Appendix. We use the language of Newman and Penrose [10]
essentially based on a differential form structure [11] throughout
the paper.

\begin{center}
{\bf 2. Changing masses of variable-charged black holes}
\end{center}
\setcounter{equation}{0}
\renewcommand{\theequation}{2.\arabic{equation}}

        In this section, by solving Einstein-Maxwell field equations with the
{\sl variable}-charge $e(r)$, we develop the relativistic aspect
of Hawking radiation in classical space-time metrics. The
calculation of Newman-Penrose (NP) spin coefficients is being
carried out through the technique developed by McIntosh and
Hickman [11] in (+,\,--,\,--,\,--) signature. In the formulation
of this relativistic aspect of Hawking radiation, we do not impose
any condition in the field equations except considering the charge
$e$ to be a function of the polar coordinate $r$. \vspace* {0.2in}

     {\sl 2.1. Variable-charged Reissner-Nordstrom solution} \\\\
We consider the non-rotating variable-charged solution with the
assumption that the charge $e$ of the body is a function of
coordinate $r$ :

\begin{equation}
ds^2=\{1-{2M\over r}+{e^2(r)\over r^2}\}\,du^2+2du\,dr-r^2(d\theta^2
+{\rm sin}^2\theta\,d\phi^2).
\end{equation}
Initially when $e$ is constant,
this metric will reduce to non-rotating charged Reissner-Nordstrom
solution. The complex null tetrad vectors for this metric are chosen as
\begin{eqnarray*}
\ell^a=\delta^a_2,\;\;\
\end{eqnarray*}
\begin{equation}
n^a=\delta^a_1-{1\over 2}\{1-{2M\over
r}+{e^2(r)\over r^2}\}\,\delta^a_2,\;\;\
\end{equation}
\begin{eqnarray*}
m^a={1\over\surd 2r}\,(\delta^a_3+{i\over {\rm sin}\theta}\,\delta^a_4),
\end{eqnarray*}
where $\ell^a$,\, $n^a$ are real null vectors and $m^a$ is complex null vector.
Using these null tetrad vectors one could calculate the spin coefficients,
Ricci scalars and Weyl
scalars as follows:
\begin{eqnarray*}
\kappa=\sigma=\nu=\lambda=\pi=\tau=\epsilon=0,\;\;\ \rho=-{1\over r},\;\
\beta=-\alpha={1\over {2\surd 2r}}\,{\rm cot}\theta,
\end{eqnarray*}
\begin{equation}
\mu=-{1\over 2\,r}\{1-{2M\over r}+{e^2(r)\over r^2}\},\;\;\
\end{equation}
\begin{eqnarray*}
\gamma={1\over 2\,r^2}\,\{M+e(r)\,e'(r)-{e^2(r)\over r^2}\},\;\
\end{eqnarray*}
\begin{eqnarray*}
\phi_{11}={1\over 4\,r^2}\,\{e'^2(r)+e(r)\,e''(r)\}-{e(r)e'(r) \over r^3}+
{e^2(r)\over 2\,r^4},\;\
\end{eqnarray*}
\begin{equation}
\Lambda = -{1\over 12\,r^2}\,\{e'^2(r)+e(r)\,e''(r)\},
\end{equation}
\begin{equation}
\psi_2={1\over 6\,r^2}\,\{e'^2(r)+e(r)\,e''(r)\}-{1\over r^3}\,\{M
+ e(r)e'(r)- {e^2(r) \over r}\},\;\
\end{equation}
where the Weyl curvature scalar $\psi_2$, Ricci scalars $\phi_{11}$ and
$\Lambda$ are defined by
\begin{eqnarray*}
\psi_2 \equiv -C_{abcd}\,\ell^a\,m^b\,\overline{m}^c\,n^d,\;\;\,
\end{eqnarray*}
\begin{equation}
\phi_{11} \equiv -{1\over 4}\,R_{ab}(\ell^a\,n^b+m^a\,\overline{m}^b),\;\;\,
\Lambda \equiv {1\over 24}\,R_{ab}\,g^{ab}.
\end{equation}
Here, a prime denotes the derivative with respect to $r$.
If the energy momentum tensor is of electromagnetic fields, then
the Ricci tensor $R_{ab}$ is proportional to the Maxwell stress tensor [8]
that is
\begin{equation}
\phi_{AB} = k\,\phi_A\,\overline \phi_B,\;\;\ k=8\pi G/c^2
\end{equation}
with $A,B$ = 0,1,2 and the NP Ricci scalar
\begin{equation}
\Lambda \equiv {1\over 24}\,R_{ab}\,g^{ab}=0.
\end{equation}
Hence, vanishing $\Lambda$ in (2.8) with (2.4) leads
\begin{equation}
e^2(r) = 2\,rm_1 + C
\end{equation}

where  $m_1$ and $C$ are real constants. Then the Ricci scalar
becomes
\begin{equation}
\phi_{11} ={C\over 2\,r^4}.
\end{equation}
Thus, the Maxwell scalar $\phi_1={1\over 2}
F_{ab}(\ell^a\,n^b+\overline{m}^a\,m^b)$ takes the form, by identifying the
real constant $C$ = $e^2$,
\begin{equation}
\phi_1={1\over\surd 2}\,e\,r^{-2},
\end{equation}
showing that the Maxwell scalar $\phi_1$ does not change its form by
considering the charge $e$ to be a function of $r$ in Einstein-Maxwell field
equations. Here, by using equation (2.9) in (2.3) and (2.5), we have the
resulting NP
quantities
\begin{equation}
\mu=-{1\over 2\,r}\{1-{2\over r}(M - m_1)+{e^2\over r^2}\},\;\;\
\end{equation}
\begin{eqnarray*}
\gamma={1\over 2\,r^2}\,\{(M - m_1)-{e^2\over r}\},\;\
\end{eqnarray*}
\begin{equation}
\psi_2=-{1\over r^3}\,\{(M - m_1)-{e^2\over r}\},\;\;\
\phi_1={1\over\surd 2}\,e\,r^{-2},
\end{equation}
and the metric (2.1) takes the form
\begin{equation}
ds^2=\{1-{2\over r}(M - m_1)+{e^2\over r^2}\}\,du^2+2du\,dr-r^2(d\theta^2
+{\rm sin}^2\theta\,d\phi^2).
\end{equation}
        This means that the mass $M$ of non-rotating black hole (2.1) is lost a
quantity $m_1$ at the end of the first electrical radiation. This
loss of mass is agreeing with Hawking's discovery that the
radiating objects must lose its mass [2]. On this losing mass,
Wald [12] has pointed that a black hole will lose its mass at the
rate as the energy is radiated. If one considers the same process
for second time taking $e$ in (2.14) to be function of $r$ with
the mass $M - m_1$ in Einstein-Maxwell field equations, then the
mass may again be decreased by another constant $m_2$ (say); that
is, after the second time radiation the total mass might become $M
- (m_1 + m_2)$. This is due to the fact, that the Maxwell scalar
$\phi_1$ with condition (2.8) does not change its form after
considering the charge $e$ to be function of $r$ for the second
time as $\Lambda$ calculated from the Einstein-Maxwell field
equations has to vanish for electromagnetic fields with $e(r)$.
Hence, if one repeats the same process for $n$-times considering
every time the charge $e$ to be function of $r$, then one would
expect the solution to change gradually and the total mass becomes
$M - (m_1 + m_2 + m_3 + . . . + m_n)$ and therefore the metric
(2.14) may take the form:
\begin{equation}
ds^2=[1-{2\over r}\,{\cal M}+{e^2\over r^2}]\,du^2
+2du\,dr-r^2(d\theta^2+{\rm sin}^2\theta\,d\phi^2),
\end{equation}
where the mass of the black hole after the radiation of $n$-times would be
\begin{equation}
{\cal M} = M - (m_1 + m_2 + m_3 + . . . + m_n).
\end{equation}
So, accordingly the changed NP quantities are
\begin{equation}
\mu=-{1\over 2\,r}\{1-{2\over r}{\cal M}+{e^2\over r^2}\},\;\;\
\end{equation}
\begin{equation}
\gamma={1\over 2\,r^2}\,\{{\cal M}-{e^2\over r}\},\;\
\end{equation}
\begin{equation}
\psi_2=-{1\over r^3}\,\{{\cal M}-{e^2\over r}\}.\;\;\
\end{equation}

This suggests that for every electrical radiation, the original mass $M$ of the
non-rotating black hole may lose some quantity. Thus, it seems reasonable
to expect that, taking Hawking's
radiation of black holes into account,
such continuously  lose of mass may lead to evaporate the
original mass $M$. In case the black hole has evaporated down to
the Planck mass, the mass $M$ may not exactly equal to the
continuously lost quantities $m_1 + m_2 + m_3 + . . . + m_n$. That
is, according to the second possibility of Hawking and Israel [5]
quoted above, there may left a small quantity of mass, say,
Planck mass of about $10^{-5}g$ with continuous electrical
radiation. Otherwise, when $M$ = $m_1 + m_2 + m_3 + . . . + m_n$
for a complete evaporation of the mass, ${\cal M}$ would be zero,
rather than, leaving behind a Planck-size mass black hole
remnant. At this stage the non-rotating black hole geometry
would have the electric charge $e$ only, but no mass, that the
line element would be of the form
\begin{equation}
ds^2=(1+{e^2\over r^2})\,du^2+2du\,dr-r^2(d\theta^2+{\rm sin}^2\theta\,d\phi^2).
\end{equation}
That is, the black hole might be radiated away  all its mass
completely just leaving the electrical radiation. Then the
classical space-time metric (2.20) may regard to represent a
non-rotating zero mass naked singularity as pointed out by Davies
[13]. Here we refer such object with zero mass as `instantaneous'
naked singularity -- a naked singularity that exists for an
instant and then continues its electrical radiation to create
negative mass. [Syteinmular, King and LoSota [14] refer a
spherically symmetric star, which radiates away all its mass as
`instantaneous' naked singularity that exists only for an instant
and then disappears. This `instantaneous' naked singularity is
also mentioned in [15]].

   The time taken between two consecutive radiations is supposed
to be very short that one may not physically realize how quickly
radiations take place. Thus it seems natural to expect the
existence of `instantaneous' naked singularity with zero mass
only for an instant before continuing its next radiation to
create negative mass naked singularity.
        It may also be possible in the reasonable theory of black holes that, as
a black hole is invisible in nature, one may not know that in the
universe, a particular black hole has mass or not, but electrical
radiation may be detected on the black hole surface. So, there may
be some radiating black holes without masses in the universe,
where the gravity may depend only on the electric charge, $i.e.$,
$\psi_2$=$e^2/r^4$, not on the mass of the black holes. Just after
the exhaustion of the mass, if one continues the remaining
solution (2.20) to radiate, there may be a formation of new mass
$m^*_1$ (say). If one repeats the electrical radiation further,
the new mass might increase gradually and then, the metric (2.20)
with the new mass would become
\begin{equation}
ds^2=(1+{2\over r}\,{\cal M}^* + {e^2\over r^2})\,du^2+2du\,dr-r^2(d\theta^2
+{\rm sin}^2\theta\,d\phi^2),
\end{equation}
where the new mass is given by
\begin{equation}
{\cal M}^*=m^*_1 + m^*_2 + m^*_3 + m^*_4 + . . .
\end{equation}
Comparing the metrics (2.15) and (2.21) one could observe that
the classical space-time (2.21) may describe a non-rotating
spherical symmetric star with a negative mass ${\cal M}^*$.
Such objects with negative masses are referred as naked
singularities [1,4,5]. The metric (2.21) may be regarded to
describe the incorporation of the third possibility of
Hawking and Israel [5] in the case of non-rotating singularity.
Here it is noted that the creation of negative mass naked
singularity is mainly based on the continuous electrical
radiation of the variable charge $e(r)$ in the energy momentum
tensor of Einstein-Maxwell equations. This also indicates the
incorporation of Boulware's suggestion [6] that `the
stress-energy tensor may be used to calculate the change in the
metric due to the radiation'.

This new mass ${\cal M}^*$ would never decrease, rather might increase gradually
as the radiation continues forever. Then the spin coefficients
for the metric (2.21) are
\begin{equation}
\mu=-{1\over 2\,r}\{1+{2\over r}{\cal M*}+{e^2\over r^2}\},\;\;\
\end{equation}
\begin{equation}
\gamma=-{1\over 2\,r^2}\,\{{\cal M*}+{e^2\over r}\},\;\
\end{equation}
\begin{equation}
\psi_2={1\over r^3}\,\{{\cal M*}+{e^2\over r}\},\;\;\
\end{equation}
and the unchanged Maxwell scalar $\phi_1$ is given in (2.11). Thus, one has seen
the changes in the mass of the non-rotating charged black hole after every
radiation. Hence, it follows the theorem cited above in the case of
non-rotating {\sl variable-charged} black hole.
\vspace* {0.2in}

     {\sl 2.2. Variable-charged Kerr-Newman solution} \\\\
Here, one may incorporate the Hawking radiation, how the rotating
variable-charged black hole affects in the classical space-time
metric when the electric charge $e$ is taken as a function of $r$
in the Einstein-Maxwell field equations. The line element with
$e(r)$ is
\begin{eqnarray}
ds^2&=&[1-R^{-2}\{2Mr-e^2(r)\}]\,du^2+2du\,dr \cr
&+&2aR^{-2}\{2Mr-e^2(r)\}\,{\rm sin}^2
\theta\,du\,d\phi-2a\,{\rm sin}^2\theta\,dr\,d\phi \cr
&-&R^2d\theta^2-\{(r^2+a^2)^2
-\Delta^*a^2\,{\rm sin}^2\theta\}\,R^{-2}{\rm sin}^2\theta\,d\phi^2,
\end{eqnarray}
where
\begin{equation}
R^2=r^2+a^2{\rm cos}^2\theta,\,\
\Delta^*=r^2-2Mr+a^2+e^2(r).
\end{equation}
This metric will also reduce to rotating Kerr-Newman solution when $e$ becomes
constant. The null tetrad vectors are chosen as
\begin{eqnarray*}
\ell^a=\delta^a_2,\;\ n^a={1\over R^2}\left[(r^2+a^2)\,\delta^a_1
-{\Delta^*\over 2}\,\delta^a_2+a\,\delta^a_4\right],
\end{eqnarray*}
\begin{equation}
m^a={1\over\surd 2R}\,\left[ia\,{\rm sin}\theta\,\delta^a_1
+\delta^a_3+{i\over{\rm sin}\theta}\,\delta^a_4\right],
\end{equation}
where $R=r+i\,a\,{\rm cos}\theta$. Then we solve the Einstein-Maxwell field
equations for the metric (2.26) and write only the changed NP quantities
\begin{equation}
\mu=-{1\over{2\overline R\,R^2}}\{r^2-2Mr+a^2+e^2(r)\},\;\;\
\end{equation}
\begin{equation}
\gamma={1\over{2\overline R\,R^2}}\,[\{r-M+e(r)\,e'(r)\}\overline
R-\{r^2-2Mr+a^2+e^2(r)\}],\;\
\end{equation}
\begin{equation}
\psi_2={1\over{\overline R\,\overline R\,R^2}}\,[-MR
+ e^2(r)-e(r)e'(r)\,{\overline R}]+{1\over{6\,R^2}}\{e'^2(r)+e(r)\,e''(r)\},\;\
\end{equation}
\begin{equation}
\phi_{11}={1\over {2\,R^2\,R^2}}\,\{e^2(r)-2r\,e(r)e'(r)\}+
{1\over 4R^2}\,\{e'^2(r)+e(r)\,e''(r)\},\;\
\end{equation}
\begin{equation}
\Lambda = -{1\over 12\,R^2}\,\{e'^2(r)+e(r)\,e''(r)\},
\end{equation}
where a prime denotes the derivative with respect to $r$.
\vspace* {0.2in}

   Now, like equation (2.8) in non-rotating black hole, the
scalar $\Lambda$ must vanish for this rotating metric. Thus,
vanishing $\Lambda$ of the equation (2.33) implies that
\begin{equation}
e^2(r) = 2\,rm_1 + C
\end{equation}
where  $m_1$ and $C$ are real constants of integration.
Then, using this result in equation (2.32) we obtain the Ricci
scalar
\begin{equation}
\phi_{11}={C\over 2\,R^2\,R^2}.
\end{equation}
Accordingly, the Maxwell scalar may become, after identifying the
constant $C=e^2$,
\begin{equation}
\phi_1={e\over\surd 2\,\overline R\,\overline R}.
\end{equation}
This is the same Maxwell scalar of the electrovac Kerr-Newman
black hole (see appendix) where the charge $e$ is constant.
   Hence, from the Einstein-Maxwell equations we have the changed NP
quantities  \\
\begin{equation}
\mu=-{1\over{2\overline
R\,R^2}}\{r^2-2r(M-m_1)+a^2+e^2\},\;\;\
\end{equation}
\begin{eqnarray*}
\gamma={1\over{2\overline
R\,R^2}}\,[\{r-(M-m_1)\}\overline R-\{r^2-2r(M-m_1)+a^2+e^2\}],\;\
\end{eqnarray*}
\begin{equation}
\psi_2={1\over{\overline R\,\overline R\,R^2}}\,\{-(M-m_1)R +
e^2\},\;\
\end{equation}

\begin{equation}
\phi_{11}={e^2\over 2\,R^2\,R^2}.
\end{equation}
Thus, the rotating solution with a new constant $m_1$
takes the following form
\begin{eqnarray}
ds^2&=&[1-R^{-2}\{2r(M-m_1)-e^2\}]\,du^2+2du\,dr \cr
&+&2aR^{-2}\{2r(M-m_1)-e^2\}\,{\rm sin}^2
\theta\,du\,d\phi-2a\,{\rm sin}^2\theta\,dr\,d\phi \cr
&-&R^2d\theta^2-\{(r^2+a^2)^2
-\Delta^*a^2\,{\rm sin}^2\theta\}\,R^{-2}{\rm sin}^2\theta\,d\phi^2,
\end{eqnarray}
where
\begin{equation}
\Delta^*=r^2-2r(M-m_1)+a^2+e^2.
\end{equation}
This introduction of constant $m_1$ in the metric (2.40) suggests
that the first electrical radiation of rotating black hole may reduce
the original gravitational mass $M$ by a quantity $m_1$. If one
considers another radiation by taking $e$ in (2.40) to be a function
of $r$ with the mass $M-m_1$, then the Einstein-Maxwell field equations yield to
reduce this mass by another constant quantity $m_2$; i.e., after the second
radiation, the mass may become $M-(m_1+m_2)$. Here again, the  Maxwell scalar
$\phi_1$ remains the same form after the second radiation also.
Thus, if one considers the $n$-time radiations taking every time the
charge $e$ to be function of $r$, the Maxwell scalar $\phi_1$ will be the
same, but the metrics may be of the following form:
\begin{eqnarray}
ds^2&=&[1-R^{-2}\{2r{\cal M}-e^2\}]\,du^2+2du\,dr \cr
&+&2aR^{-2}\{2r{\cal M}-e^2\}\,{\rm sin}^2
\theta\,du\,d\phi-2a\,{\rm sin}^2\theta\,dr\,d\phi \cr
&-&R^2d\theta^2-\{(r^2+a^2)^2
-\Delta^*a^2\,{\rm sin}^2\theta\}\,R^{-2}{\rm sin}^2\theta\,d\phi^2,
\end{eqnarray}
where the total mass of the black hole, after the $n$-time
radiations may take the form
\begin{equation}
{\cal M}=M-(m_1 + m_2 + m_3 + m_4 + . . .+ m_n).
\end{equation}
Taking Hawking's radiation of black holes, one might expect that
the total mass of  black hole may be
radiated away just leaving ${\cal M}$ equivalent to Planck mass
of about
$10^{-5}g$, that is, $M$ may not be exactly equal to
$m_1 + m_2 + m_3 + m_4 + . . .+ m_n$, but has a difference of
about Planck-size mass, as in the case of
non-rotating black hole. Otherwise, the total mass of black hole
may be evaporated completely after continuous radiation when
${\cal M} = 0$, that is,
$M = m_1 + m_2 + m_3 + m_4 + . . .+ m_n$.
Here one may regard that the rotating variable-charged black
hole might be radiated completely away all its mass just leaving the
electrical charge $e$ only. One could observe this situation in
the form of classical space-time metric as
\begin{eqnarray}
ds^2&=&(1+e^2\,R^{-2})\,du^2+2du\,dr \cr
&+&2ae^2\,R^{-2}\,{\rm sin}^2
\theta\,du\,d\phi-2a\,{\rm sin}^2\theta\,dr\,d\phi \cr
&-&R^2d\theta^2-\{(r^2+a^2)^2
-\Delta^*a^2\,{\rm sin}^2\theta\}\,R^{-2}{\rm sin}^2\theta\,d\phi^2,
\end{eqnarray}
with the charge $e$, but no mass,
where $\Delta^*=r^2+a^2+e^2$.
The metric (2.44) may describe a rotating `instantaneous' naked
singularity with zero mass. At this stage, the Weyl scalar
$\psi_2$ takes the form
\begin{equation}
\psi_2={e^2\over{\overline R\,\overline R\,R^2}}\;\
\end{equation}
showing the gravity on the surface of the remaining solution
depending only on the electric charge $e$; however, the Maxwell
scalar $\phi_1$ remains the same as in (2.36).
For future use, we mention the
changed NP spin coefficients
\begin{equation}
\mu=-{1\over{2\overline R\,R^2}}\{r^2+a^2+e^2\},\;\;\
\end{equation}
\begin{eqnarray*}
\gamma={1\over{2\overline R\,R^2}}\,[r\overline R-\{r^2+a^2+e^2\}].\;\
\end{eqnarray*}

It suggests that there may be rotating black holes in the universe
whose masses are completely radiated; their gravity depend only on
the electric charge of the body and their metrics look like the
one given in the equation (2.44). It appears that the idea of this
evaporation of masses of radiating black holes may be agreed with
that of Hawking's evaporation of black holes. Unruh [16] has
examined various aspects of black hole evaporation based on
Schwarzschild metric. It is worth studying the nature of such
black holes (2.44) or in the case of non-rotating (2.20). This
might give a different nature, which one has not yet come across
so far in the reasonable theory of black holes. \vspace* {0.2in}

   Here, one may consider again the charge $e$ to be function
of radial coordinate $r$ for next radiation in (2.44), so that one
must get from the Einstein's field equations the scalar
$\Lambda$ as given in equation (2.33). Then the vanishing of this
$\Lambda$ for electromagnetic field, there may be
creation of a new mass (say $m^*_1$) in the remaining space-time geometry.
If this radiation process continues forever, the new mass may
increase gradually as
\begin{equation}
{\cal M}^*=m^*_1 + m^*_2 + m^*_3 + m^*_4 + . . .  .\,\,
\end{equation}
However, it appears that this new mass would never decrease. Then the
space-time geometry may take the form
\begin{eqnarray}
ds^2&=&[1+R^{-2}\{2r{\cal M}^*+e^2\}]\,du^2+2du\,dr \cr
&+&2aR^{-2}\{2r{\cal M}^*+e^2\}\,{\rm sin}^2
\theta\,du\,d\phi-2a\,{\rm sin}^2\theta\,dr\,d\phi \cr
&-&R^2d\theta^2-\{(r^2+a^2)^2
-\Delta^*a^2\,{\rm sin}^2\theta\}\,R^{-2}{\rm sin}^2\theta\,d\phi^2,
\end{eqnarray}
where
\begin{equation}
\Delta^*=r^2+2r{\cal M}^*+a^2+e^2.
\end{equation}
The Weyl scalar $\psi_2$ and other NP coefficients are calculated from the
Einstein-Maxwell field equations as
\begin{equation}
\psi_2={1\over{\overline R\,\overline R\,R^2}}\,\{{\cal
M}^*R+e^2\}.\;\
\end{equation}
\begin{equation}
\mu=-{1\over{2\overline R\,R^2}}\{r^2+2r{\cal
M}^*+a^2+e^2\},\;\;\
\end{equation}
\begin{eqnarray*}
\gamma={1\over{2\overline
R\,R^2}}\,[\{(r+{\cal M}^*)\overline
R-\{r^2+2r{\cal M}^*+a^2+e^2\}],\;\
\end{eqnarray*}
with $\phi_1$ given in (2.36). The metric (2.48) may be regarded
to describe a rotating negative mass naked singularity. We have
presented the possible changes in the mass of the rotating charged
black hole without affecting the Maxwell scalar $\phi_1$ and
accordingly, metrics are cited for future use. Thus, this
completes the proofs of other parts of the theorems for the
rotating charged black hole. \vspace* {0.2in}

\begin{center}
{\bf 3. Conclusion}
\end{center}
\setcounter{equation}{0}
\renewcommand{\theequation}{3.\arabic{equation}}

        In the above section, it has shown the changes in the mass of charged
black holes after every radiation. However, according to Hawking evaporation
rate [2], the quantities $m_1's$ to be reduced from the mass of the black hole,
may not be equal i.\,e., $m_1 \neq m_2 \neq m_3 \neq . . . .  \,\,.$
It appears from the results discussed in this paper that the
black hole radiation process is mainly based on the electrical
radiation of the variable-charged $e(r)$ in the energy momentum
tensor describing the change in mass in classical space-time
metrics. The creation of negative mass naked singularities is
also due to the continuous electrical radiation. This clearly
suggests that an electrical radiating black hole, non-rotating or
rotating would not disappear completely, rather, would form a
negative mass naked singularity with regards to the nature of
rotation. However, the disappearance of such black hole during
the radiation process may occur only for an instant, just at the
time of formation of `instantaneous' naked singularity with zero
mass. The formation of naked singularity of negative mass is also
Hawking's suggestion mentioned in the introduction above [4].
This suggests that, if one excepts the continuous electrical
radiation to lead the complete evaporation of the original mass
of black holes, then the same radiation might also lead the
creation of new mass to form negative mass naked singularities.
The classical space-time metrics, rotating or non-rotating
discussed above, would describe the possible life style of
electrically radiating black holes during their radiation
process.

Also, we know from the above that the change in the mass of black
holes takes place due to the Maxwell scalar ${\phi_1}$, remaining
unchanged in the field equations. So, if the Maxwell scalar
${\phi_1}$ is absent from the space-time geometry, there will be
no radiation and consequently, no reduction in the mass of the
black hole. Therefore, one could not expect, theoretically  to
observe  such `{\sl relativistic change}' in the mass in the case
of uncharged Schwarzschild as well as Kerr black holes. Here, a
brief description of the metrics formed after radiation is being
presented.

        Under the transformation [1]
\begin{equation}
du=dt-{(r^2+a^2)\over \Delta^*}\,dr,\;\;\:
 d\phi'=d\phi-{a\over\Delta^*}\,dr,
\end{equation}
the metric (2.44) is written in the Boyer-Lindquist coordinates
($t,r,\theta,\phi$)
\begin{eqnarray}
ds^2&=&(1+e^2\,R^{-2})\,dt^2-{R^2\over\Delta^*}dr^2
-R^2d\theta^2 \cr
&-&\{(r^2+a^2)^2
-\Delta^*a^2\,{\rm sin}^2\theta\}\,R^{-2}{\rm
sin}^2\theta\,d\phi^2  \cr
&-&2ae^2\,R^{-2}\,{\rm sin}^2\theta\,dt\,d\phi.
\end{eqnarray}
Clearly, one might see that the equation
\begin{equation}
\Delta^*\equiv r^2+a^2+e^2=0
\end{equation}
has no solution for the real radial coordinate $r$ unless
$a^2+e^2< 0$. Hence, the metric (3.2)
has no event horizon and may not describe a black hole for $a^2+e^2> 0$. Also
there is no stationary limit for this metric as $g_{tt}>0$
has no real solution for $r$.

The metric (2.48) could be written in the Boyer-Lindquist coordinates
($t,r,\theta,\phi$)
\begin{eqnarray}
ds^2&=&[1+R^{-2}\{2r\,{\cal M}^*+e^2\}\,]\,dt^2-{R^2\over\Delta^*}dr^2
-R^2d\theta^2 \cr
&-&\{(r^2+a^2)^2
-\Delta^*a^2\,{\rm sin}^2\theta\}\,R^{-2}{\rm
sin}^2\theta\,d\phi^2 \cr
&-&2a(2r\,{\cal M}^*+e^2)\,R^{-2}\,{\rm sin}^2\theta\,dt\,d\phi.
\end{eqnarray}
where
\begin{equation}
\Delta^* = r^2+2r\,{\cal M}^*+a^2+e^2.
\end{equation}
If one expects the rotating charged metric (3.4) to be a black
hole, then one might find $r_{\pm}=-{\cal M}^*\pm \surd {({\cal
M}^{*2}-a^2-e^2)}$ as the roots of the equation $\Delta^*=0$. It
appears that the event horizon at $r_+$ for this metric would be
very small comparative to the mass ${\cal M}^*$ given in (2.47).
Also the stationary limit at $r_e =-{\cal M}^*\pm\surd {({\cal
M}^{*2}-a^2{\rm cos}^2\theta-e^2)}$ for the surface $g_{tt}>0$ is
also quite small. So the true singularity at $R^2=r^2+a^2{\rm
cos}^2\theta = 0$ may be the only singularity of the coordinate
components (3.4). Thus, it seems reasonable to conclude that the
metric (3.4) might describe a negative mass naked singularity.
However, this solution certainly characterizes a rotating
electrovac Petrov type $D$ space-time as $\psi_2\neq 0$, with the
charge $e$.

        The metric (2.48) can be written in the coordinates $(t,\,x,\,y,\,z)$
\begin{eqnarray}
ds^2&=&dt^2-dx^2-dy^2-dz^2 \cr
      &+&{(2r\,{\cal M}^*+e^2)\,r^2 \over(r^4+a^2\,z^2)}\,[dt-
      {1\over(r^2+a^2)}\,\{r(xdx+ydy)+a(xdy-ydx)\}-{1\over r}\,zdz]^2
\end{eqnarray}
where $r$ is defined, with a sign difference in terms of $(x,\,y,\,z)$ [1]
\begin{equation}
r^4-(x^2+y^2+z^2-a^2)\,r^2-a^2\,z^2=0,\,
\end{equation}
and the $(x,\,y,\,z)$ have the following relations
\begin{eqnarray*}
x=(r\,{\rm cos}\phi + a\,{\rm sin}\phi)\,{\rm sin}\theta,\;\;\;  y=(r\,{\rm
sin}\phi - a\,{\rm cos}\phi)\,{\rm sin}\theta,
\end{eqnarray*}
\begin{equation}
z=r\,{\rm cos}\theta,\;\;\; x^2+y^2=(r^2+a^2)\,{\rm sin}^2\theta.
\end{equation}
Then, the above metric (3.6) is the Kerr-Schild form with
\begin{equation}
g_{ab} = \eta_{ab} + 2\,H(x,\,y,\,z)\, L_a\,L_b,
\end{equation}
where $\eta_{ab}$ is the flat metric and
\begin{equation}
H(x,\,y,\,z)={(2r\,{\cal M}^*+e^2)\,r^2 \over 2\,(r^4+a^2\,z^2)}
\end{equation}
\begin{equation}
L_a\,dx^a=dt-
      {1\over(r^2+a^2)}\,\{r(xdx+ydy)+a(xdy-ydx)\}-{1\over r}\,zdz.
\end{equation}
From the Kerr-Schild form metric (3.6), one clearly sees that when ${\cal
M}^*=0$, it reduces to the Kerr-Schild form of the metric (3.2) and it again
comes to flat metric with $e=0$. Here, one may note the difference from the
Kerr-Newman metric [1] that there is a change in sign before the square
parenthesis of (3.6) for the metric (2.48) and also for the metric (2.44) with
${\cal M}^* = 0$

        As the metric (2.48) or (3.4) describes an electrovac Petrov
type $D$ space-time ($\psi_2 \neq 0$), it may be better to mention that the spin
coefficients
of the metric (2.48) satisfy {\sl Chandrasekhar's relation}. That, Chandrasekhar
[17] has established a relation of spin coefficients $\rho$,
$\mu$,
$\tau$,
$\pi$
in the case of an affinely parameterized geodesic vector, generating an integral
which is constant along the geodesic in a {\sl vacuum} Petrov type $D$
space-time
\begin{equation}
{\rho\over \overline\rho}={\mu\over\overline\mu}={\tau\over\overline\pi}
={\pi\over\overline\tau}.
\end{equation}
The original derivation of this relation is purely based on the vacuum Petrov
type $D$ space-time with $\psi_2\neq 0$, $\psi_0=\psi_1=\psi_3=\psi_4=0$ and
$\phi_{01}=\phi_{02}=\phi_{10}
=\phi_{20}=\phi_{12}=\phi_{21}=\phi_{00}=\phi_{22}=\phi_{11}=\Lambda=0$.
However, after the introduction of Killing-Yano scalar $\chi_1$ in the above
relation as [18]
\begin{equation}
{\rho\over \overline\rho}={\mu\over\overline\mu}={\tau\over\overline\pi}
={\pi\over\overline\tau}=-{\overline\chi_1\over\chi_1},
\end{equation}
where $\chi_1={1\over 2} f_{ab}(\ell^a\,n^b+\overline{m}^a\,m^b)$ with
Killing-Yano tensor $f_{ab}$ satisfying the Killing-Yano equations
\begin{equation}
f_{ab;c}+f_{ac;b}=0,
\end{equation}
it is found that Chandrasekhar's relation holds true for {\sl
non-vacuum} Petrov type $D$ space-times. The importance of KY
tensor in General Relativity seems to lie on Carter's remarkable
result [19] that the separation constant of Hamilton-Jacobi
equation (for charged orbits) in the Kerr space-time gives a
fourth constant. In fact, this constant arises from the scalar
field $K_{ab}v^a\,v^b$ which has vanishing divergent along a unit
vector $v^a$ tangent to an orbit of charged particle. Here $K_{ab}
=f_{ma}f^m_b$. For examples, (i) in Kerr-Newman solution where the
source of gravitational field is electromagnetic field, the
relation (3.13) takes
\begin{equation}
{\rho\over \overline\rho}={\mu\over\overline\mu}={\tau\over\overline\pi}
={\pi\over\overline\tau}=-{\overline\chi_1\over\chi_1}
={R\over\overline R},
\end{equation}
where $R=r+i\,a\,{\rm cos}\theta$ and the spin coefficients $\rho$, $\mu$,
$\tau$, $\pi$ are given in Appendix.. The Killing-Yano scalar $\chi_1$ is
obtained as [18]
\begin{equation}
\chi_1=iC(r-i\,a\,{\rm cos}\theta)
\end{equation}
where $C$ is a real constant. Here the Killing-Yano tensor  for Kerr-Newman
metric is
\begin{equation}
f_{ab}=4a\,{\rm cos}\theta\,n_{[a}\ell_{b]}+4i\,r\,m_{[a}\overline m_{b]}.
\end{equation}
and  accordingly the Killing tensor $K_{ab}=f_{ma}\,f^m_b$
\begin{equation}
K_{ab}=-8\{\,a^2{\rm cos}^2\theta\,\ell_{(a}n_{b)}+r^2\,m_{(a}\overline
m_{b)}\}.
\end{equation}

(ii) in Kantowski-Sachs metric, where the source of gravitation is dust
i.e. $T_{ab}$=${\rho}^*u_a\,u_b$ with ${\rho}^*$ being the density of the
material dust particles,
$u_a$ the unit time like vector, the relation (3.13) becomes [18]
\begin{equation}
{\rho\over \overline\rho}={\mu\over\overline\mu}=-{\overline\chi_1\over\chi_1}
=1.
\end{equation}
Here, the Killing-Yano scalar is found as $\chi_1={1\over 2}\,iC\,Y(t)$, where
$C$ is constant and $Y(t)$ is the function appeared in the Kantowski-Sachs
metric. Then the Killing-Yano tensor and the Killing tensor for this metric take
the form
\begin{equation}
f_{ab}=2\,iY(t)\,m_{[a}\overline m_{b]}\;\;\; and\;\;\;
K_{ab}=2\,Y^2(t)\,m_{(a}\overline m_{b)}.
\end{equation}
From the above examples, it seems reasonable to refer the relation
(3.12) as {\sl `Chandrasekhar's identity'} as mentioned by Fernandes and Lun
[20]. Thus, it concludes that the metric (2.48) admits the relation
(3.13) with the same $\chi_1$ as in (3.16).

        Taking Hawking radiation into account, one might mention here that the
time taken of one radiation to another would be very short that
one may not, practically realize after losing $m_1$ from the mass
how quickly $m_2$ is being reduced  and so on, as seen above in
Section 2. It appears that the metric (2.20) or (2.44) without
mass may occur {\sl only} for a very short period, as the
radiation continues further. Thus, we have incorporated Hawking
radiation in relativistic viewpoint in curved space-time geometry.
One may expect that the metrics with mass ${\cal M}^*$ in
equations (2.21) and (2.48) might have different nature from
Reissner-Nordstrom as well as Kerr-Newman solutions.

\begin{appendix}
\setcounter{equation}{0}
\renewcommand{\theequation}{A\arabic{equation}}
\begin{center}
{\bf Appendix}
\end{center}

     {\bf (a) Reissner-Nordstrom solution}: This is a spherically
symmetric solution
\begin{equation}
ds^2=(1-{2M\over r}+{e^2\over r^2})\,du^2+2du\,dr-r^2(d\theta^2
+{\rm sin}^2\theta\,d\phi^2),
\end{equation}
where $M$ and $e$ are the mass and charge respectively. For this
metric one chooses the null tetrad vectors
\begin{equation}
\ell^a=\delta^a_2,\;\;\ n^a=\delta^a_1-{1\over 2}(1-{2M\over
r}+{e^2\over r^2})\,\delta^a_2,\;\;\
m^a={1\over\surd 2r}\,(\delta^a_3+{i\over {\rm sin}\theta}\,\delta^a_4).
\end{equation}
The NP quantities are
\begin{eqnarray*}
\kappa=\sigma=\nu=\lambda=\pi=\tau=\epsilon=0,\;\;\ \rho=-{1\over r},\;\
\beta=-\alpha={1\over {2\surd 2r}}\,{\rm cot}\theta,
\end{eqnarray*}
\begin{equation}
\mu=-{1\over 2\,r}(1-{2M\over r}+{e^2\over r^2}),\;\;\
\gamma={1\over 2\,r^3}\,(M\,r-e^2),\;\ \psi_2=-(M\,r-e^2)\,r^{-4},\;\
\phi_1={1\over\surd 2}\,e\,r^{-2}.
\end{equation}
\vspace* {0.2in}

        {\bf (b) Kerr-Newman solution}: The line element is
\begin{eqnarray}
ds^2&=&\{1-R^{-2}(2Mr-e^2)\}\,du^2+2du\,dr+2aR^{-2}(2Mr-e^2)\,{\rm sin}^2
\theta\,du\,d\phi-2a\,{\rm sin}^2\theta\,dr\,d\phi \cr
&-&R^2d\theta^2-\{(r^2+a^2)^2
-\Delta^*a^2\,{\rm sin}^2\theta\}\,R^{-2}{\rm sin}^2\theta\,d\phi^2,
\end{eqnarray}
where $R^2=r^2+a^2{\rm cos}^2\theta$,  $\Delta^*=r^2-2Mr+a^2+e^2$.
The null tetrad vectors are
\begin{eqnarray*}
\ell^a=\delta^a_2,\;\ n^a={1\over R^2}\left[(r^2+a^2)\,\delta^a_1
-{\Delta^*\over 2}\,\delta^a_2+a\,\delta^a_4\right],
\end{eqnarray*}
\begin{equation}
m^a={1\over\surd 2R}\,\left[ia\,{\rm sin}\theta\,\delta^a_1
+\delta^a_3+{i\over{\rm sin}\theta}\,\delta^a_4\right].
\end{equation}
where $R=r+i\,a\,{\rm cos}\theta$. Then the NP quantities are
\begin{eqnarray*}
\kappa=\sigma=\nu=\lambda=\epsilon=0,\cr
\rho=-{1\over\overline R},\;\;\;\;
\mu=-{\Delta^*\over{2\overline R\,R^2}},
\end{eqnarray*}
\begin{eqnarray*}
\alpha={(2ai-R\,{\rm cos}\theta)\over{2\surd 2\overline R\,\overline R\,{\rm
sin}\theta}},\;\;\;
     \beta={{\rm cot}\theta\over2\surd 2R},
\end{eqnarray*}
\begin{equation}
\pi={i\,a\,{\rm sin}\theta\over{\surd 2\overline R\,\overline R}},\;\;\;
\tau=-{i\,a\,{\rm sin}\theta\over{\surd 2R^2}},
\end{equation}
\begin{eqnarray*}
\gamma={1\over{2\overline R\,R^2}}\,\left[(r-M)\overline
R-\Delta^*\right],
\end{eqnarray*}
\begin{eqnarray*}
\phi_0=\phi_2=0,\;\;\ \phi_1={e\over{\surd 2\overline R\,\overline R}},
\end{eqnarray*}
\begin{equation}
\psi_0=\psi_1=\psi_3=\psi_4=0,\;\;
\psi_2=-{M\over{\overline R^3}}+{e^2\over{R\,\overline R^3}}.
\end{equation}
From the above quantities one can deduce (a) Kerr vacuum solution
($e=0, a\ne 0$), (b) Reissner-Nordstrom solution ($e\ne 0, a=0$) and
(c) Schwarzschild solution ($e=a=0$).
\end{appendix}
\vspace* {0.2in}

        {\bf Acknowledgement}: The author acknowledges his appreciation for
hospitality received from Inter-University Centre for Astronomy and
Astrophysics (IUCAA), Pune, India during the preparation of this paper.
I am thankful the unknown referees for remarkable suggestions
which lead to significant improvement of an earlier version.


\begin{thebibliography}{99}

\bibitem{x}Chandrasekhar S 1983 {\sl The Mathematical Theory of
Black Holes}  Clarendon Press, Oxford.
\bibitem{x}Hawking S W 1974 {\sl Nature} {\bf 248} 30; 1975 {\sl
Commun. math. Phys.} {\bf 43} 199.
\bibitem{x} Zhang Z 1991 Physics of Black Holes : Classical, Quantum and
Astrophysical in {\sl Black Hole Physics} eds. V De Sabbata and Z Zhang, Kluwer
Academic Publishers, Dordrecht.
\bibitem{x}Hawking S W 1976 {\sl Phys. Rev. D} {\bf 14} 2460
\bibitem{x}Hawking S W and Israel W 1980 ``An Introductory
survey"
in {\sl General Relativity: An Einstein Centenary Survey} eds.
by S W Hawking and W Israel, Cambridge University Press.
\bibitem{x}Boulware D G 1976 {\sl Phys. Rev. D} {\bf 13} 2169.
\bibitem{x}Vaidya P C 1951 {\sl Proc. Indian Acad. Sci.} {\bf A33} 264,
Reprinted
1999 {\sl Gen. Rel. Grav.} {\bf 31} 119.
\bibitem{x}Hartle J B and Hawking S W 1976 {\sl Phys. Rev. D} {\bf 13}
2188.
\bibitem{x}Bekenstein J D 1973 {\sl Phys. Rev. D} {\bf 7} 2333;
Nieuwenhuizen T M 1998 {\sl Phys. Rev. D} {\bf 81} 2201; Sivaram C
2000 {\sl Phys. Rev. Lett.} {\bf 84} 3209; 2001 {\sl Gen. Rel. Grav.}
{\bf 33} 175, MacGibbon J H 1991 {\sl Phys. Rev. D} {\bf 44} 376.
\bibitem{x}Newman E T and Penrose R 1962 {\sl J. Math. Phys.} {\bf 3} 566.
\bibitem{x}McIntosh C B G  and Hickman  M S 1985 {\sl Gen. Rel. Grav.} {\bf
17} 111; and Debever R, McLeneghan R G and Tariq N 1979 {\sl Gen.
Rel. Grav.} {\bf 10} 853.
\bibitem{x}Wald R M 1975 {\sl Commun. Math.Phys.} {\bf 45} 9; 1991
``Black holes and Thermodynamics" in {\sl Black Hole
Physics} eds. V De Sabbata and Z Zhang Kluwer Academic Publishers, Dordrecht.
\bibitem{x}Davies P C W 1980 ``Quantum field in Curved Space" in
{\sl General Relativity and Gravitation}, Vol 2. ed by A Held,
Plenum Press, New York.
\bibitem{x} Steinmular B, King A R and LoSota J P 1975 {\sl Phys
Lett} {\bf 51A} 191
\bibitem{x}Tipler F J, Clerke C J S and Ellis G F R 1980
``Singularities and Horizons -- A review article" in {\sl
General Relativity and Gravitation: One Hundred Year After the birth of
Albert Einstein}, Vol 2, ed A Held, Plenum Press, New York.
\bibitem{x}Unruh W G 1976 {\sl Phys. Rev. D} {\bf 14} 870.
\bibitem{x} Reference [1] p 324.
\bibitem{x}Ibohal Ng 1997 {\sl Astrophys and SpaceSc} {\bf 249} 73.
\bibitem{x} Carter B 1968 {\sl Phys. Rev.} {\bf 174} 1559 and
1968 {\sl Commun. Math. Phys.} {\bf 10} 280; 1973 Black Hole Equilibrium
States in C Dewitt and B C Dewitt eds. {\sl Black Holes} Gordon and
Breach Sci. Publ. New York.
\bibitem{x}Fernandes J F Q and Lun A W C 1997 {\sl J. Math. Phys.} {\bf
38} 330.

\end{thebibliography}
\end{document}